\documentclass[12pt]{amsart}  
\usepackage{amscd} 
\usepackage{mathrsfs}
\newtheorem{theorem}{Theorem}[section]

\newtheorem{corollary}[theorem]{Corollary} 
\theoremstyle{definition}

\newtheorem{example}[theorem]{Example}
\newtheorem{remark}[theorem]{Remark} 
\numberwithin{equation}{section}

\def\H{\mathscr H}  
\def\K{\mathscr K}  
\def\M{\mathscr M} 
\def\N{\mathscr N} 
\def\D{\mathscr D} 
\def\RE{\mathbb R} 
\def\CO{{\mathbb C}}
\def\Ker{\mathscr K} 
\def\Ran{\mathscr R}
\def\Im{\text{\rm Im}\,} 
\def\uno{\mathsf 1}
\def\zero{\mathsf 0}
\def\fh{\mathfrak h}
\def\ff{\mathfrak f}
\def\E{\mathsf {E}}
\def\P{{\mathsf P}}

\def\p{\par\noindent}

\def\trip{\left\{\fh,\beta_1,\beta_2\right\}}
\def\e{\mathsf e}

\newsymbol\subsetneq 2328
                                                                    
\begin{document}

\title[Self-adjoint Extensions of Restrictions] {Self-adjoint
Extensions of Restrictions}

\author{Andrea Posilicano}

\address{Dipartimento di Scienze Fisiche e Matematiche,  Universit\`a
dell'Insubria, I-22100 Como, Italy}

\email{posilicano@uninsubria.it}

\thanks{
(primary), 
47B38, 35J25 (secondary)}

\begin{abstract}
We provide a simple recipe for obtaining all 
self-adjoint extensions, together with their resolvent,
of the symmetric operator $S$ obtained by restricting the self-adjoint
operator $A:\D(A)\subseteq\H\to\H$ to the dense, closed with respect
to the graph norm,
subspace $\N\subset \D(A)$. Neither the knowledge of $S^*$ nor 
of the deficiency spaces of $S$ is required. Typically $A$ is
a differential operator
and $\N$  is the kernel of some trace (restriction) operator 
along a null subset. We parametrise the extensions by the bundle
$p:\E(\fh)\to\P(\fh)$,
where $\P(\fh)$ denotes the set of orthogonal
projections in the Hilbert space $\fh\simeq
\D(A)/\N$ and $p^{-1}(\Pi)$ is the set of self-adjoint
operators in the range of $\Pi$. 
The set of self-adjoint operators in $\fh$, i.e. 
$p^{-1}(\uno)$,  
parametrises the relatively prime extensions. Any
$(\Pi,\Theta)\in \E(\fh)$ determines a boundary condition 
in the domain of the corresponding extension
$A_{\Pi,\Theta}$ and explicitly appears in the formula for the resolvent
$(-A_{\Pi,\Theta}+z)^{-1}$. The connection
with both von 
Neumann's and Boundary Triples theories of self-adjoint
extensions is explained. Some examples related to quantum graphs, 
to Schr\"odinger operators with  
point interactions and to elliptic boundary
value problems are given.
\end{abstract}

{\maketitle }
\begin{section}{Introduction.}
On the Hilbert
space $\H$ with scalar product $\langle\cdot,\cdot\rangle$ we consider the 
self-adjoint operator $$A:\D(A)\subseteq\H\to\H\,.$$  
We denote by $\H_A$ the Hilbert space
given by the operator domain $\D(A)$ endowed
with the graph inner product
$$\langle\phi,\psi\rangle_A:=\langle A\phi,A\psi\rangle+\langle\phi,\psi\rangle
\,.
$$
Given a closed subspace $\N\subset\H_A$ which is dense in $\H$, we denote by 
$S$ the closed, densely defined, symmetric operator obtained by restricting 
$A$ to $\N$. Here our aim is to find all self-adjoint extensions of
$S$ and to provide their resolvent.
\par  
Since $\N$ is closed we have $\H_A=\N\oplus\N^{\perp}$ and thus
$\N$ coincides with the kernel of the orthogonal projection onto 
$\N^{\perp}$. 
Since $\N^{\perp}\simeq \H_A/\N$ is a Hilbert space, 
without loss of generality we can suppose that $\N$ coincides with the
kernel of a surjective bounded linear operator
$$
\tau:\H_A\to\fh 
$$
with $\fh$ a Hilbert space. This choice 
has some advantages in practical applications, where usually $\tau$ is
given in advance. Indeed typically $A$ is a differential operator and $\tau$ 
is some trace (restriction) operator along a null subset.  \par
In Section 2, by using the results in \cite{[P1]}, 
we construct, by explicitly giving their
resolvents (see Theorem \ref{estensioni}), a family of self-adjoint
extensions of the symmetric
operator $S$. Such extensions are parametrised by couples
$(\Pi,\Theta)$, where $\Pi$ is a orthogonal projection in $\fh$ and
$\Theta$ is a self-adjoint operator in the range of $\Pi$. The
resolvent Kre\u \i n-like formula (\ref{krein}) we provide (see
\cite{[K1]}, \cite{[K2]}, \cite{[S]} for the
original Kre\u \i n's formula; also see \cite{[GMT]} and
references therein) resembles the one
obtained, by Boundary Triple theory, in \cite{[MM]}, Corollary 5.6, 
for the case of a dual pair of operators, and in \cite{[AP]}, \cite{[Pa]} 
for the case of a single symmetric
operator. There a resolvent formula is given in terms of a couple 
$(B_1,B_2)$ of bounded linear operators which satisfy a 
commutativity hypothesis and a non-degeneracy one 
(see (\ref{comm}) and (\ref{nondeg}) in Remark 4.4), while here 
we impose no further conditions on the couple
$(\Pi,\Theta)$.\par
Then, by using the
results in \cite{[P2]}, we give (see Theorem \ref{star}) 
an alternative description which shows 
how the couple $(\Pi,\Theta)$ induces a boundary condition 
in the operator domain of the corresponding
extension. Again this has connections with 
\cite{[MM]}, \cite{[AP]} and \cite{[Pa]}. We conclude Section 2 by giving (see
Theorem \ref{additive}) an additive representation for the self-adjoint 
extensions of the symmetric operator $S$. The analogous result  
in the case of relatively prime extensions was obtained
in \cite{[P2]}.\par
In Section 3 we explore the connection with von Neumann's theory of
self-adjoint extensions \cite{[N]}. By extending the results 
in \cite{[P2]}, Section 4, 
we explicitly provide (see Theorem \ref{vonNeumann}) a
bijection from unitary operators $U:\K_+\to \K_-$ ($\K_\pm$
denoting the defect spaces of $S$) to couples $(\Pi,\Theta)$ in such a
way that $A_U=A_{\Pi,\Theta}$, where $A_U$ is the extension given by
von Neumann's Theory and $A_{\Pi,\Theta}$ is the extension given in
Theorem \ref{estensioni}. This show, as a byproduct, that our
construction provides all the self-adjoint extensions of $S$ (see
Corollary \ref{corollario}). Thus the whole set of 
self-adjoint extensions of the symmetric
operator $S$ is parametrised by the bundle
$p:\E(\fh)\to\P(\fh)$,
where $\P(\fh)$ denotes the set of orthogonal
projections in $\fh$ and $p^{-1}(\Pi)$ is the set of self-adjoint
operators in the range of $\Pi$. This kind of parametrisation is
compatible with the one obtained, in the case $A$ is injective with 
a bounded inverse, in
\cite{[G1]}, Theorem II 2.1 (also see \cite{[V]}, Section 1). We refer
to Example 5.5 in Section 5 below for
more details in the case of applications to elliptic boundary value problems. 
\par
In Section 4 we explore, by using the results in \cite{[P3]}, 
the connection with Boundary Triples theory of 
self-adjoint extensions (see e.g. \cite{[B]}, \cite{[Koch]}, \cite{[GG]}, 
\cite{[DM1]}, \cite{[BGP]}). 
In particular we recover Theorem 5.3 in
\cite{[A]} by which any self-adjoint relation in
$\fh\oplus\fh$ is of the kind 
${\mathscr  G}(\Theta)\oplus\fh_0^{\perp}$, 
where $\fh_0\subseteq\fh$ 
is a closed subspace, $\Theta$ is some self-adjoint operator in
$\fh_0$ and ${\mathscr  G}(\Theta)$ denotes its graph. The
connection with different parametrisations of the set
of self-adjoint relations is explicitly
given in Theorem \ref{s.a.relations}. This provides the bridge
between our Kre\u \i n-like formula (\ref{krein}) and the one given in
\cite{[MM]}, \cite{[AP]} and \cite{[Pa]}.  
\par
Finally, in Section 5, we give some applications by 
examples related to quantum graphs, to Schr\"odinger operators with  
point interactions and to boundary
value problems for the Laplace operator on bounded domains.

\end{section}
\begin{section}{Self-adjoint Extensions.}

In the following, given a linear operator $L$ we denote by 
$$\D(L)\,,\quad
\Ker(L)\,,\quad\Ran(L)\,,\quad\rho(L)$$ its domain, kernel, range and
resolvent set respectively. \par
Let 
$$A:\D(A)\subseteq\H\to\H\,,$$ 
$$
\tau:\H_A\to\fh\,,\qquad
\Ran(\tau)=\fh\,,\quad
\overline{\Ker(\tau)}=\H\,,
$$
and
$$
S:\Ker(\tau)\subseteq\H\to\H
$$
be the respectively self-adjoint, bounded and symmetric operators
considered in the introduction. \par 
For any $z$ in $\rho(A)$ the linear operator
$$R(z):=(-A+z)^{-1}$$ is bounded on $\H$ onto $\H_A$. 
Thus for any $z\in\rho(A)$  we can define the bounded operator
$$
G(z):=(\tau R(\bar z))^* :\fh\to\H
$$
(here the $*$ denotes the Hilbert space adjoint). 
The surjectivity of $\tau$ makes $G(z)$ injective 
. 
By \cite{[P2]}, Lemma 2.1, one has that, given the surjectivity
hypothesis $\Ran(\tau)=\fh$, the density one 
$\overline{\Ker(\tau)}=\H$ is equivalent to 
\begin{equation}\label{inj}
\Ran(G(z))\cap \D(A)=\left\{0\right\}\,.
\end{equation}
By the first resolvent identity one easily obtains (see
\cite{[P1]}, Lemma 2.1), for any $z$ and $w$ in $\rho(A)$, 
\begin{equation}\label{resg}
(z-w)\,R(w)G(z)
=G(w)- G(z)\,,
\end{equation}
thus 
\begin{equation}\label{reg}
\Ran (G(w)-G(z))\in \D(A)\,.
\end{equation}
Let $\Gamma(z):\fh\to\fh $, $z\in\rho(A)$, be a family of bounded 
linear operators such that 
\begin{equation}\label{gamma}
\Gamma(z)-\Gamma(w)=
(z-w)\,G(\bar w)^*G(z)
\end{equation} 
and
\begin{equation}\label{gamma*}
\Gamma(z)^*=\Gamma(\bar z)\,.
\end{equation}
The class of such families is not void. Indeed by (\ref{resg}) and
the definition of $\Gamma(z)$ it is easy to check (see \cite{[P1]},
Lemma 2.2, for the details) that any of such a family differs by a 
$z$-independent bounded self-adjoint operator
from the family
$\hat \Gamma_{(w)}(z)$ defined by
\begin{equation}\label{gammahat}
\hat \Gamma_{(w)}(z):=\,\tau\left(\frac{G(w)+G(\bar
w)}{2}-G(z)\right)\,,\quad w\in\rho(A)\,.
\end{equation}
Note that $\hat\Gamma_{(w)}(z)$ is
well defined by (\ref{reg}). \par Given the orthogonal projection $$
\Pi:\fh\to\fh\,,
$$ 
pose
$$
\fh_0:=\Ran(\Pi) 
$$
and let
$$
\Theta:\D(\Theta)\subseteq\fh_0\to\fh_0
$$
be self-adjoint. Then we define the closed operator
$$
\Gamma_{\Pi,\Theta}(z):=(
\Theta +\Pi\,\Gamma(z)\Pi) :\D(\Theta)\subseteq \fh_0\to\fh_0\,,
$$
and the open set
$$
Z_{\Pi,\Theta}:=\left\{z\in\rho(A)\,:\,0\in\rho
(\Gamma_{\Pi,\Theta}(z))\right\}\,.
$$
The next theorem is nothing but the combination of Theorem 2.1 and
Proposition 2.1 in \cite{[P1]} when the bounded and surjective
linear map there denoted by $\tau$ is given by
$\Pi\tau:\H_A\to\fh_0$. We give a short
self-contained proof for the reader's convenience.
\begin{theorem} \label{estensioni} Let $A$, $\tau$, $S$, $\Pi$, $\Theta$ and 
$\Gamma_{\Pi,\Theta}$ be as above. Then $$\CO\backslash\RE\subseteq 
Z_{\Pi,\Theta}$$ and the bounded linear operator 
\begin{equation}\label{krein}
R_{\Pi,\Theta}(z):=R(z)+
G(z)\Pi\,\Gamma_{\Pi,\Theta}(z)^{-1}\Pi G(\bar z)^*\,,
\qquad z\in Z_{\Pi,\Theta}\,,
\end{equation}
is a resolvent of the self-adjoint extension 
$A_{\Pi,\Theta}$ of $S$ defined by
$$
A_{\Pi,\Theta}:\D(A_{\Pi,\Theta})
\subseteq \H\to\H\,,\qquad (-A_{\Pi,\Theta}+z)\phi:=(-A+z)\phi_z\,,
$$
$$
\D(A_{\Pi,\Theta}):=\left\{\,\phi\in\H\, :\, \phi=
\phi_z+G(z)\Pi\,\Gamma_{\Pi,\Theta}(z)^{-1}\Pi\tau\,\phi_z,
\,\phi_z\in \D(A)\,\right\}\,.
$$
Such a definition is $z$-independent and the 
decomposition of $\phi$ appearing 
in $\D(A_{\Pi,\Theta})$ is unique. 
\end{theorem}
\begin{proof} By (\ref{gamma}) and (\ref{gamma*}), denoting by 
$(\cdot,\cdot)$ the scalar product in $\fh$ and by $\|\cdot\|$ the
  norm in $\H$, one has
$$
|(\zeta_0,\Gamma_{\Pi,\Theta}(z)\zeta_0)|^2\ge \Im(z)^2\|G(z)\zeta_0\|^4
$$
for any $\zeta_0\in\fh_0$. 
Thus $\Gamma_{\Pi,\Theta}(z)$ is injective for all
$z\in\CO\backslash\RE$ by the injectivity of $G(z)$.
Since $$
\Ran(\Gamma_{\Pi,\Theta}(z))^{\perp}
=\Ker(\Gamma_{\Pi,\Theta}(z)^*)=
\Ker(\Gamma_{\Pi,\Theta}(\bar z))=\{0\}\,,$$
 the range of $\Gamma_{\Pi,\Theta}(z)$
is dense. 
Since $G(z)\Pi=\left(\Pi\tau R(\bar z)\right)^*:\fh_0\to\H$, 
the range of 
$G(z)\Pi$ is closed by the surjectivity of both $\Pi$, $\tau$, $R(\bar
z)$ and by the closed range theorem. Thus
$$
\gamma:=\inf_{\zeta_0\in\fh_0\backslash\{0\}}\frac{\|G(z)\zeta_0\|}
{\|\zeta_0\|}>0\,.
$$
Therefore
$$
\inf_{\zeta_0\in\fh_0\backslash\{0\}}\frac{\|\Gamma_{\Pi,\Theta}(z)\zeta_0\|}
{\|\zeta_0\|}\ge\inf_{\zeta_0\in\fh_0\backslash\{0\}}
\frac{|(\zeta_0,\Gamma_{\Pi,\Theta}(z)\zeta_0)|}{\|\zeta_0\|^2}
\ge |\Im(z)|\,\gamma^2>0
$$
and the range of
$\Gamma_{\Pi,\Theta}(z)$ is closed. Thus for any
$z\in\CO\backslash\RE$ the closed operator 
$\Gamma_{\Pi,\Theta}(z)$ 
is both injective and surjective. By the inverse mapping theorem 
$\CO\backslash\RE\subseteq 
Z_{\Pi,\Theta}$.\par

By using (\ref{gamma}), a simple computation 
(see \cite{[P1]}, page 115) shows that   
$R_{\Pi,\Theta}(z)$
satisfies the resolvent identity 
\begin{equation}\label{res}
(z-w)\,R_{\Pi,\Theta}(w)R_{\Pi,\Theta}(z)=
R_{\Pi,\Theta}(w)-R_{\Pi,\Theta}(z)
\end{equation}
and, by (\ref{gamma*}), 
\begin{equation}\label{symm}
R_{\Pi,\Theta}(z)^*=R_{\Pi,\Theta}(\bar z)\,.
\end{equation} 
Moreover, by (\ref{inj}), 
$R_{\Pi,\Theta}(z)$ is injective. Thus 
$$A_{\Pi,\Theta}:=z-R_{\Pi,\Theta}(z)^{-1}$$ is well-defined on 
$$\D(A_{\Pi,\Theta}):=\Ran (R_{\Pi,\Theta}(z))\,,$$ 
is $z$-independent by (\ref{res}) and is
symmetric by (\ref{symm}). 
It is self-adjoint since $\Ran(-A_{\Pi,\Theta}\pm i)=\H$ by 
construction.
\end{proof}
\begin{remark} By the successive results in Section 4 and 
by \cite{[DM1]}, Propositions 1 and 2, Section 2, one
  has that $$\lambda\in
  \sigma_p(A_{\Pi,\Theta})\cap\rho(A)\quad\iff\quad 
0\in\sigma_p(\Gamma_{\Pi,\Theta}(\lambda))\,,$$ where
  $\sigma_p(\cdot)$ denotes point spectrum. An analogous result holds
  for the continuous spectrum. Regarding the
  eventual eigenvectors and their multiplicity, by \cite{[P3]}, Theorem 3.4, 
  one has that 
$$ 
G(\lambda):\Ker(\Gamma_{\Pi,\Theta}(\lambda))\to 
\Ker(-A_{\Pi,\Theta}+\lambda)$$
is a bijection for any $\lambda\in \sigma_p(A_{\Pi,\Theta})\cap\rho(A)$.
\end{remark}
Now we provide an alternative description of the self-adjoint
extensions obtained in the previous theorem. This result will
show how the couple $(\Pi,\Theta)$ induces a boundary conditions on
the elements in the operator domain of the corresponding extension.
\par  
Since $$
\Gamma(z)-\hat\Gamma(z)=\hat\Theta\,,\quad\hat\Gamma(z):=\hat\Gamma_{(i)}(z)\,,
$$
where $\hat\Theta$ is a $z$-independent bounded symmetric operator
in $\fh$, one has
$$\Gamma_{\Pi,\Theta}(z)=(\Theta+\Pi\hat\Theta \Pi)+\Pi\,\hat
\Gamma(z)\Pi\,.$$
Since $\Theta$ is an arbitrary self-adjoint operator in $\fh_0$, 
from  now on we will take, without loss
of generality,
\begin{equation}
\Gamma_{\Pi,\Theta}(z)=\hat\Gamma_{\Pi,\Theta}(z)
:=\Theta+\Pi\,\hat\Gamma(z)\Pi
\,.
\end{equation}
Let us define
$$
R\equiv R_+:=R(i)\,,\quad R_-:=R(-i)\,,
$$
$$
G_+:=G(i)\,,\quad G\equiv G_-:=G(-i)\,,
$$
$$
G_*:=\frac{1}{2}\,\left(G_-+G_+\right)\,.
$$
Thus
$$
\hat\Gamma(z)=\tau(G_*-G(z))\,.
$$
\begin{remark}
The choice $w=i$ in the above definitions 
is not essential. Any different $w\in\CO\backslash\RE$ would lead 
to analogous results. Whereas the
behaviour of a single extension depends on the choice of the family 
$\Gamma(z)$, and hence depends on the choice of $w$, the whole family
of extensions does not. In any case one can easily connect any two 
parametrisations provided by different families $\Gamma_{1}(z)$ and 
$\Gamma_{2}(z)$: it suffices to use the substitution 
$$\Theta\leftrightarrow \Theta+\Pi(\Gamma_1(z)-\Gamma_2(z))\Pi\,.$$ 
By choosing $w\in\CO\backslash\RE$ (in particular $w=i$) we
can treat the case of an arbitrary self-adjoint extension without making $w$
$Z_{\Pi,\Theta}$-dependent. In 
the case one works with a specific operator, different (more appealing)
choices are possible. Indeed we can interpret Section 5 below (where $A$ is the Laplacian) as a proof, by
examples, of a different version of next Theorem 
\ref{star} valid in
the case $w=0$. In examples 5.1, 5.2 and 5.5 one has 
$w=0\in\rho(A)$ whereas in examples 5.3 and 5.4 $w=0\in
\sigma(A)\backslash
\sigma_{p}(A)$, thus showing that there are situations in which 
$w$ it not even required
to be in $\rho(A)$ (see \cite{[P4]} for a study of the case $w=0\in
\sigma(A)\backslash
\sigma_{p}(A)$ in a general setting. By Remark 3.3 in \cite{[P5]} the
hypotheses on $\tau$ required in \cite{[P4]} can be relaxed, thus
allowing wider applications).  
\end{remark}
The next theorem is nothing but  Theorem in
\cite{[P2]} (also see \cite{[P4]}, Corollary 3.2) when one uses the bounded and surjective map (there
denoted by $\tau$) given by
$\Pi\tau:\H_A\to\fh_0$ and notes that, by (\ref{resg}), 
$$
G_*=-iRG+G\,.
$$ 
Note that in the definition of $\D(A)$ now appears the boundary 
condition
$$
\Pi\tau
\phi_*=\Theta\zeta_\phi\,.
$$
\begin{theorem}\label{star} The self-adjoint extension
$A_{\Pi,\Theta}$ can be re-written as
$$
A_{\Pi,\Theta}:\D(A_{\Pi,\Theta})\subseteq\H\to\H\,,\qquad
A_{\Pi,\Theta}\phi=A\phi_*+RG\zeta_\phi\,,
$$
\begin{align*}
&\D(A_{\Pi,\Theta})
=\left\{\phi\in\H\,:\,\phi=\phi_*+G_*\zeta_\phi\,,\right.\\ 
&\left.\phi_*\in \D(A)\,,\, \zeta_\phi\in
\D(\Theta)\,,\,\Pi\tau
\phi_*=\Theta\zeta_\phi
\right\}\,.
\end{align*}
\end{theorem}
We conclude this section by giving 
an additive representation for the self-adjoint extensions of the
symmetric operator $S$. Let us denote by $\H_A^*$ the Hilbert space
obtained by completing $\H$ with respect to the scalar product
$$
[\phi,\psi]_A:=\langle\phi,(A^2+1)^{-1}\psi\rangle\equiv\langle R\phi,R\psi\rangle\,.
$$
Let $$
\bar A:\H\to\H_A^*
$$
be the bounded extension of 
$$
A:\H_A\subseteq\H\to\H_A^*
$$
and let us denote by $$
(\cdot,\cdot)_A:\H_A\times\H_A^*\to\CO
$$
the pairing obtained by extending the scalar product in $\H$
$$
\langle\cdot,\cdot\rangle:\H_A\times\H\subseteq \H_A\times\H_A^*\to\CO\,.
$$
We define then
$$
\tau^*:\fh\to\H_A^*
$$
by
$$
\forall\,\phi\in\H_A\,,\qquad (\phi,\tau^*\zeta)_A=(\tau\phi,\zeta)\,,
$$
where $(\cdot,\cdot)$ denotes the scalar product in $\fh$.\par
The next theorem is nothing but Theorem 3.4 in
\cite{[P2]} when one uses the bounded and surjective map (there
denoted by $\tau$) given by
$\Pi\tau:\H_A\to\fh_0$.
\begin{theorem}\label{additive} When restricted to
  $\D(A_{\Pi,\Theta})$ the 
linear operator
$$
\tilde A:\D(\tilde A)\subseteq\H\to\H_A^*\,,\quad\tilde A\phi:=\bar
A\phi+\tau^*
\zeta_\phi\,,
$$
$$
\D(\tilde A)
=\left\{\phi\in\H\,:\,\phi=\phi_*+G_*\zeta_\phi\,,\, 
\phi_*\in \D(A)\,,\, \zeta_\phi\in\fh\right\}\,,
$$
is $\H$-valued and coincides with $A_{\Pi,\Theta}$. 
\end{theorem} 
\end{section}
\begin{section}{The connection with von Neumann's Theory.}
In this section we explore the connection between the
results given in Section 2 and von Neumann's theory of
self-adjoint extensions. As a byproduct we will obtain that our
construction provides all self-adjoint extensions of the symmetric
operator $S$.\par
By defining the deficiency spaces
$$
\K_\pm:=\Ker\left(-S^*\pm i\right)
$$
and posing
$$
\N:=\Ker (\tau)\,,
$$
von Neumann's theory says that 
$$
\D(S^*)= \N\oplus
\K_+\oplus \K_-\,,\quad
S^*(\phi_0+\phi_++\phi_-)=A\phi_0+i\phi_+-i\phi_-\,,
$$ 
the direct sum decomposition being orthogonal with
respect to the graph inner product of $S^*$; any self-adjoint
extension $A_U$ of $S$ is then obtained
by restricting $S^*$ to a subspace of the kind $\N\oplus {\mathscr G}(U)
$, where
$U:\K_+\to \K_-$ is unitary and ${\mathscr G}(U)$ denotes its graph. 
\par 
In the next theorem we pose
$\hat\Gamma:=\hat\Gamma(i)\equiv\frac{1}{2}\,
\tau(G_--G_+)$, thus $\hat\Gamma^*=-\hat\Gamma$. Moreover in 1) we
use the decomposition $\fh=\fh_0\oplus\fh_0^\perp$.
\begin{theorem} \label{vonNeumann} 
1) The linear operators 
$$G_\pm:\fh\to
\K_\pm$$ are continuous bijection and the linear operator 
$$U:\K_+\to \K_-\,,\qquad
U:=-\,G_-((\uno+2(\Theta-\Pi\,\hat\Gamma \Pi)^{-1}\Pi\,
\hat\Gamma\Pi)\oplus\uno)\, G_+^{-1}\,,
$$
which can be alternatively
re-written, in the case $\Theta$ is bounded, as 
$$
U=-\,G_-((\Theta-\Pi\,\hat\Gamma \Pi)^{-1}
(\Theta+\Pi\,\hat\Gamma \Pi)\oplus\uno)\, G_+^{-1}\,,
$$
is unitary. 
The corresponding extension $A_U$ given by von Neumann's theory
coincides with the self-adjoint operator $A_{\Pi,\Theta}$.\p
2) Let $A_U$ be the self-adjoint extension of $S$
corresponding, by von Neumann's theory, to the unitary operator
$U:\K_+\to \K_-$ and let $\M\subseteq\H_A$ be the closed subspace
$\M:=\D(A_U)\cap\D(A)$. 
Then there exists a closed subspace $\fh_0$ (see (\ref{h0}) for the
precise definition),  
$$
\fh_0\subseteq\left(\tau\left[\M\cap\N^\perp\right]\right)^\perp\subseteq\fh\,,
$$
such that, denoting by $\Pi$ the orthogonal projection onto 
$\fh_0$ and by $
U_A:=\left(-A+i\right)\left(-A-i\right)^{-1} 
$ the Cayley transform of $A$, the linear operator 
$\Theta:\D(\Theta)
\subseteq\fh_0\to\fh_0$
defined by
$$\Theta:=i\Pi G^*(U-U_A)(U+U_A)^{-1}G\Pi
$$
is self-adjoint. 
The corresponding self-adjoint extension $A_{\Pi,\Theta}$ 
coincides with $A_U$.
\end{theorem}
\begin{proof} 1) The first half of the theorem is consequence of
  Theorem 4.1 in \cite{[P2]} 
when one uses the bounded and surjective map (there denoted by $\tau$) 
given by $\Pi\tau:\H_A\to\fh_0$.\p
2) Since $\N\subseteq\M$ we have the orthogonal decomposition
$$
\H_A=\N\oplus(\M\cap\N^\perp)\oplus\M^\perp\,,
$$
so that, if $\phi=\phi_0+\phi_1+\phi_2$,                                       
$$
\tau\phi=\tau\phi_1+\tau\phi_2\,.
$$
If $\tau\phi_1=\tau\phi_2$ then $\phi_1-\phi_2\in\N$ which gives 
$\phi_1=\phi_2=0$ since both $\phi_1$ and $\phi_2$ are in $\N^\perp$
and $\phi_1\perp\phi_2$.
Thus 
$$
\fh=\tau[\M\cap \N^\perp]+\tau[\M^\perp]\,,
\qquad \tau[\M\cap \N^\perp]\cap\tau[\M^\perp]
=\{0\}\,. 
$$
Since $\tau$ is continuous and surjective, by the open
mapping theorem
both $\tau[\M\cap \N^\perp]$ and $\tau[\M^\perp]$ are closed and therefore
there exists a unique continuous projection $P$ in $\fh$ such
that 
$$
\Ker(P)=\fh_1:=  \tau[\M\cap \N^\perp]\,,
\qquad \Ran (P)=\fh_2:=\tau[\M^\perp]\,.
$$
Moreover
$$
P\tau:\H_A\to\fh_2
$$
is a continuous surjection with $\Ker(P\tau)=\M$ and we can use
the results in \cite{[P2]} when the map there denoted
by $\tau$ is given by $P\tau$. In particular
by Theorem 4.3 in \cite{[P2]} the linear operator in $\fh_2$
$$\Sigma:=iP G^*(U-U_A)(U+U_A)^{-1}GP^*\,,
$$
is densely defined and self-adjoint. Moreover
$$
\hat A:\D(\hat A)\subseteq\H\to\H\,,\quad
\hat A\phi=
A\phi_*+RGP^*\xi_{\phi}\,,
$$
\begin{align*}
&\D(\hat A)
=\left\{\phi\in\H\,:\,\phi=\phi_*+G_* P^*\xi_{\phi}
\,,\right.\\
&\left.\, \phi_*\in \D(A)\,,\ \xi_\phi\in \D(\Sigma)\,,\,
 P\tau\phi_*=\Sigma\xi_{\phi}
\right\}
\end{align*}
is self-adjoint and coincides with $A_U$. Since $P^*$ is the unique 
continuous
projection in $\fh$ such that 
$$
\Ker(P^*)=\fh_2^\perp\,,\qquad\Ran(P^*)=\fh_1^\perp\,,
$$
denoting by $\Pi_2$ the orthogonal projection onto $\fh_2$,
the linear map
\begin{equation}\label{h0}
Q:=P^*\Pi_2:\fh_2\to\fh_0:=\Ran(P^*\Pi_2)
\end{equation}
is a continuous bijection. Thus
for any $\xi_\phi\in\fh_2$ there exists
an unique $\zeta_\phi\in\fh_0$ such that
$P^*\xi_\phi=\zeta_\phi$ and
$$P\tau\phi_*=\Sigma\xi_{\phi}\qquad\iff\qquad 
(Q^*)^{-1}P\tau\phi_*
=(Q^*)^{-1}\Sigma Q^{-1}\zeta_\phi\,$$ Thus $\hat A\equiv 
A_{\Pi,\Theta}$ with $$\Pi:=(Q^*)^{-1}P\,,\qquad 
\Theta:=(Q^*)^{-1}\Sigma Q^{-1}\,.$$
\end{proof}
The previous theorem shows that the self-adjoint extensions we provided
in Theorem \ref{estensioni} exhaust the set of all self-adjoint
extensions 
of $S$ (the case of
relatively prime extensions was already contained in \cite{[P2]}). Thus
we have the following
\begin{corollary}\label{corollario} The set of operators provided by 
Theorem \ref{estensioni} 
coincides with the set of all self-adjoint extensions of the symmetric
operator $S$.  Such a set is parametrised by the bundle
$p:\E(\fh)\to\P(\fh)$,
where $\P(\fh)$ denotes the set of orthogonal
projections in $\fh$ and $p^{-1}(\Pi)$ is the set of self-adjoint
operators in the range of $\Pi$. 
The set of self-adjoint operators in $\fh$, i.e. $p^{-1}(\uno)$, 
parametrises the extensions for which
$\D(A_{\uno,\Theta})\cap \D(A)=\N$ i.e. parametrises all 
relatively prime extensions of $S$.
\end{corollary}
\end{section}

\begin{section}{The connection with Boundary Triples Theory}
In this section we explore the connection between the
results given in Section 2 and Boundary Triples Theory.\par
A triple 
$\trip$, where $\fh$ is a Hilbert space with inner product 
$(\cdot,\cdot)$ and 
$$\beta_1\, :\, \D(S^*)\to\fh\,,\quad\beta_2\, :\, \D(S^*)\to\fh\,,$$ 
are two linear surjective maps, is said to be a boundary triple for $S^*$ if
$$
\langle \phi,S^*\psi\rangle-\langle S^*\phi,\psi\rangle=
(\beta_1\,\phi,\beta_2\,\psi)-
(\beta_2\,\phi,\beta_1\,\psi)\,.
$$ 
A closed subspace $\Lambda\subset \fh\oplus\fh$ is said to be a symmetric
closed relation if 
$$
\forall\,\left((\zeta_1,\zeta_2),(\xi_1,\xi_2)\right)\in\Lambda\oplus\Lambda\,,
\qquad(\zeta_1,\xi_2)=
(\zeta_2,\xi_1 )\,.
$$
Then $\Lambda$ is said to be a self-adjoint relation if it is maximal
symmetric, i.e. if it does not exists a closed symmetric relation
$\tilde \Lambda$ such that $\Lambda\subsetneq\tilde\Lambda$. Of course the
graph of a self-adjoint operator is a particular case of self-adjoint
relation. \par 
One of the main results of boundary triples theory (see e.g. 
\cite{[GG]}, Theorem 1.6, Chapter 3) is the following
\begin{theorem}\label{triples} The self-adjoint 
extensions of $S$ are parametrised by the set of self-adjoint relations
in $\fh\oplus\fh$. Any self-adjoint extension of $S$ is obtained
by restricting 
$S^*$ to the subspace 
$$\{\phi\in \D(S^*) : (\beta_1\phi,\beta_2\phi)\in\Lambda\}\,,$$ 
where $\Lambda$ is some self-adjoint relation and $\trip$ is a 
boundary triple for $S^*$. \end{theorem}
Now let us take $A$, $\tau$ and $S$ as in section 1. 
In \cite{[P3]}, Theorem 3.1, the following result  
(with slight different notations and definitions) was obtained:
\begin{theorem} The adjoint of $S$ is given by
$$
S^*:\D(S^*)\subseteq\H\to\H\,,\qquad S^*\phi=A\phi_*+RG\zeta_\phi\,,
$$
$$
D(S^*)=\left\{\,\phi\in\H\,:\,\phi=\phi_*+G_* \zeta_\phi,\ \phi_*\in
D(A),\ \zeta_\phi\in\fh\,\right\}\,.
$$
The triple $\{\,\fh,\hat\beta_1,\hat\beta_2\,\}$, where
\begin{align*}
&\hat\beta_1:\D(S^*)\to\fh\,,\qquad\hat\beta_1\,\phi:=\zeta_\phi
\,,\\
&\hat\beta_2:\D(S^*)\to\fh\,,\qquad\hat\beta_2\,\phi:=\tau\phi_*\,,
\end{align*}
is a boundary triple 
for $S^*$. 
\end{theorem}
Thus by Corollary \ref{corollario} and Theorem \ref{triples} 
we have the following 
\begin{theorem} 1) Any self-adjoint relation in $\fh\oplus\fh$ can
  be written as  
$$
\Lambda_{(\Pi,\Theta)}:=\left\{(\zeta_1,\zeta_2)\in \fh\oplus\fh\,
:\,\zeta_1\in\D(\Theta)\,,\quad
\Theta\zeta_1=\Pi\zeta_2\right\},$$
for some $(\Pi,\Theta)\in\E(\fh)$.\p 
2) Any self-adjoint extension of
$S$ is given by restricting $S^*$
to a subspace of the kind
$$
\left\{\phi\in\D(S^*)\,:\,\hat\beta_1\phi\in\D(\Theta)\,,\quad
\Theta\hat\beta_1\phi=\Pi\hat\beta_2\phi\right\}
$$
for some $(\Pi,\Theta)\in\E(\fh)$.
\end{theorem}
\begin{remark} \label{remark} By $$\fh=\Ran(\Pi)\oplus\Ker(\Pi)\,,$$ by
  $$\fh\oplus\fh\simeq\Ran(\Pi)\oplus\Ran(\Pi)
\oplus\Ker(\Pi)\oplus\Ker(\Pi)\,,$$
$$
  \Ker(\Pi)\simeq\Ker(\Pi)\oplus\{0\}\subset\fh\oplus\fh$$ 
and denoting
  by $${\mathscr G}(\Theta)\subset\Ran(\Pi)\oplus\Ran(\Pi)$$ 
the graph of $\Theta$, Theorem 4.3 gives
$$
\Lambda_{(\Pi,\Theta)}\simeq{\mathscr G}(\Theta)\oplus\Ker(\Pi)\,.
$$
This reproduces Theorem 5.3 in \cite{[A]}.\par
By \cite{[Pa]}, Proposition 4 and Lemma 5, any self-adjoint relation in
  $\fh\oplus\fh$ can be written as 
$$
\Lambda^{(B_1,B_2)}:=\left\{(\zeta_1,\zeta_2)\,:\,B_1\zeta_1=B_2\zeta_2\right\}
=\left\{(B_2^*\zeta,B_1^*\zeta)\,,\ \zeta\in\fh\right\}\,,
$$ 
where $B_1$ and $B_2$ are bounded linear operators in $\fh$ such that
\begin{equation}\label{comm}
B_1B_2^*=B_2B_1^*
\end{equation}
and
\begin{equation}\label{nondeg}
0\in\rho\left(M^{(B_1,B_2)}\right)\,,\quad M^{(B_1,B_2)}:=
\left(\begin{matrix}B_1&-B_2\\
B_2&B_1
\end{matrix}\right)\,.
\end{equation}
In the case $\fh$ is finite dimensional the condition (\ref{nondeg}) 
is equivalent either to
\begin{equation}\label{rofe1}
\Ker(B_1^*)\cap\Ker(B_2^*)
=\{0\}\end{equation}
 or to 
\begin{equation}\label{rofe2}
\det(B_1B_1^*+B_2B_2^*)\not=0\,.
\end{equation}
\end{remark}
Conditions (\ref{comm}) and (\ref{rofe2}) were obtained in
\cite{[RB]}. Their infinite dimensional analogue, as 
other equivalent conditions, are given in \cite{[DHMdS]}, Section 3.2.\par 
The connection between the representation of self-adjoint relations
in terms of $\Lambda_{(\Pi,\Theta)}$ and the one in terms of
$\Lambda^{(B_1,B_2)}$ is provided by
the following
\begin{theorem} \label{s.a.relations} Given $(\Pi,\Theta)\in\E(\fh)$
  and posing $\fh=\Ran(\Pi)\oplus\Ker(\Pi)$,
  let us define
\begin{equation}\label{AB1}
B_1:=\Theta (-\Theta+i)^{-1}\oplus\uno\,,\quad 
B_2:=(-\Theta+i)^{-1}\oplus \zero\,.
\end{equation}
Conversely, given $(B_1,B_2)$ satisfying (\ref{comm}) and
(\ref{nondeg}), let $\Pi$ be the
orthogonal projection onto $\Ker(B_2)^{\perp}$ and let $\Theta$ be the
self-adjoint operator in $\Ker(B_2)^{\perp}$ defined by
\begin{equation}\label{AB2}
\Theta:\Ran(B_2^*)\subseteq 
\Ker(B_2)^{\perp}\to\Ker(B_2)^{\perp}\,,\quad \Theta:=
\Pi B_1^*(B_2^{*}\tilde \Pi)^{-1}\Pi\,,
\end{equation}
where $\tilde \Pi$ is the orthogonal projection onto
$\Ker(B_2^*)^{\perp}$.\par
Then
$$
\Lambda_{(\Pi,\Theta)}=\Lambda^{(B_1,B_2)}\,.
$$
\end{theorem}
\begin{proof} Checking that $\Lambda^{(B_1,B_2)}\simeq
{\mathscr G}(\Theta)\oplus\Ker(\Pi)$, where $(B_1,B_2)$ is
defined by (\ref{AB1}) is straightforward.\par Conversely, by 
$$
\fh=\Ker(B_2)^{\perp}\oplus
\Ker(B_2)=\Ker(B_2^*)^{\perp}\oplus\Ker(B_2^*)\,,
$$ 
 one has
\begin{align*}
&\left\{(B_2^*\zeta,B_1^*\zeta)\,,\ \zeta\in\fh\right\}\\
=&
\left\{(\zeta_0, B_1^*( B_2^{*}\tilde \Pi)^{-1}\zeta_0+B_1^*\zeta_1)
\,,\ \zeta_0\in\Ran(B_2^*)\subseteq\Ker(B_2)^{\perp}\,,
\, \zeta_1\in \Ker(B_2^*)\right\}\,.
\end{align*}
By (\ref{comm}) $$
\zeta_1\in \Ker(B_2^*)\quad\Longrightarrow\quad B_1^*\zeta_1\in
\Ker(B_1)\,.$$
Thus
$$
\Lambda^{(B_1,B_2)}\simeq{\mathscr G}(\Theta)\oplus\Ran((\uno-\Pi) B_1^*)\,.
$$
By (\ref{nondeg}) one has 
$$
\forall\,\zeta_0\in\Ker(B_2)\quad \exists
(\zeta_1,\zeta_2)\in\fh\oplus\fh\quad \text{\rm s.t.}\quad 
(\uno-\Pi)(B_1^*\zeta_1+B_2^*\zeta_2)=\zeta_0\,.
$$
Since $\Ran(B_2^*)\subseteq \Ker(B_2)^{\perp}$, one obtains 
$$\Ran((\uno-\Pi) B_1^*)=\Ker(B_2)$$
and the proof is done.
\end{proof}
\end{section}

\begin{section}{Examples.}
For the sake of simplicity in the next examples we take $A$ equal to
the Laplace operator. With some more effort these examples could be
extended to the case in which $A$ is a variable-coefficients
differential operator. Moreover Theorem \ref{estensioni} 
could be applied to not
semi-bounded self-adjoint operators (including the case $\sigma(A)=\RE$) 
of the kind $A=iW$ where the
skew-adjoint $W$ is associated to some abstract wave equations (see
\cite {[P5]}; also see \cite{[CFP]} for an application to acoustics).
\begin{example} (The Laplacian on a bounded interval) Let 
$$A:\D(A)\subseteq L^2(0,a)\to
L^2(0,a)\,,\qquad A\psi=\psi''\,,$$ 
$$\D(A)=\{\psi\in H^2(0,a)\,:\,\psi(0+)=\psi(a-)=0\}\,,$$
$$
\tau:H^2(0,a)\to\CO^2\,,\qquad \tau\psi=\left(\psi'(0+),-\psi'(a-)\right)\,.
$$
Here $H^2(0,a)\subset C^1(0,a)$ 
denotes the usual Sobolev-Hilbert space of square
integrable functions with square integrable second order 
(distributional) derivative. 
We look for all self-adjoint extensions of the
symmetric operator 
$$S:\D(S)\subseteq L^2(0,a)\to
L^2(0,a)\,,\qquad S\psi=\psi''\,,$$
\begin{align*}
&\D(S)\equiv H^2_0(0,a)\\
:=&\left\{\psi\in
H^2(0,a)\,:\,\psi(0+)=\psi'(0+)=\psi(a-)=\psi'(a-)=0\right\}\,.
\end{align*}
Since
\begin{align*}
&\left(-\frac{d^2}{dx^2}+z\right)^{-1}\psi(x)=
\frac{\sin(\sqrt {-z}\,(a-x))}{\sqrt {-z}\sin(\sqrt {-z}\,a)}
\int_0^x{\sin(\sqrt {-z}\,y)}\,\psi(y)\,dy\\
&+
\frac{\sin(\sqrt {-z}\,x)}{\sqrt {-z}\sin(\sqrt {-z}\,a)}
\int_x^a{\sin(\sqrt {-z}\,(a-y))}\,\psi(y)\,dy\,,
\qquad z\not=-\left(\frac{n\pi}{a}\right)^2\,,
\end{align*}
\begin{align*}
&\left(-\frac{d^2}{dx^2}\right)^{-1}\psi(x)=
\frac{a-x}{a}\int_0^x y\,\psi(y)\,dy+
\frac{x}{a}\int_x^a(a-y)\,\psi(y)\,dy\,,
\end{align*}
one has
\begin{equation}\label{GZ}
G(z):\CO^2\to L^2(0,a)\,,
\end{equation}
\begin{equation*}
[G(z)\zeta](x)=
\begin{cases} \frac{\sin(\sqrt
    {-z}\,(a-x))}{\sin(\sqrt {-z}\,a)}\,\zeta_1+
\frac{\sin(\sqrt{-z}\,x)}{\sin(\sqrt {-z}\,a)}\,\zeta_2 
&\ z\not=-\left(\frac{n\pi}{a}\right)^2\\
\frac{a-x}{a}\,\zeta_1+\frac{x}{a}
\,\zeta_2 &\ z=0\,,
\end{cases}
\end{equation*}
where $\zeta\equiv(\zeta_1,\zeta_2)$ and
\begin{equation}\label{GZ*}
G(\bar z)^*:L^2(0,a)\to\CO^2\,,
\quad G(\bar z)^*\equiv\left(G(\bar z)^*_1,G(\bar z)^*_2\right)
\end{equation}
\begin{equation*}
G(\bar z)^*_1\psi=
\begin{cases} \int_0^a\frac{\sin(\sqrt
    {-z}\,(a-x))}{\sin(\sqrt {-z}\,a)}\,\psi(x)\,dx &\ z\not=-\left(\frac{n\pi}{a}\right)^2\\
\int_0^a\frac{a-x}{a}\,\psi(x)\,dx &\ z=0\,.
\end{cases}
\end{equation*}
\begin{equation*}
G(\bar z)^*_2\psi=
\begin{cases} \int_0^a\frac{\sin(\sqrt {-z}\,x)}{\sin(\sqrt {-z}\,a)}\,
\psi(x)\,dx &\ z\not=-\left(\frac{n\pi}{a}\right)^2\\
\int_0^a
\frac{x}{a}\,\psi(x)\,dx &\ z=0\,.
\end{cases}
\end{equation*}
Note that $G(z)\zeta$ solves the Dirichlet boundary value problem
\begin{align*}
&(G(z)\zeta)''=zG(z)\zeta\,,\\
&\rho\, G(z)\zeta=\zeta\,,
\end{align*}
where
$$
\rho:H^2(0,a)\to\CO^2\,,\qquad \rho\psi:=\left(\psi(0+),\psi(a-)\right)\,.
$$
Thus $$\Ran(G(z))\cap\D(A)=\{0\}\,.$$ Then one defines 
$$\Gamma(z):\CO^2\to\CO^2\,,\qquad \Gamma(z):=-\tau\, G(z)\,,$$
i.e
\begin{equation}\label{gammaz}
\Gamma(z)=\frac{\sqrt {-z}}{\sin(\sqrt{- z}\,a)}
\left(\begin{matrix}\cos(\sqrt {-z}\,a)&-1\\
-1&\cos(\sqrt {-z}\,a)\,
\end{matrix}\right)\,,\quad z\not=-\left(\frac{n\pi}{a}\right)^2\,,
\end{equation}
\begin{equation}\label{gamma0}
\Gamma(0)=\frac{1}{a}\left(\begin{matrix}{\ \ }1&-1\\
-1&{\ \ }1
\end{matrix}\right)\,.
\end{equation}
It satisfies (\ref{gamma}) and (\ref{gamma*})
by 
$$\Gamma(z)=-\tau\, G(0)+\tau(G(0)-G(z))= 
\frac{1}{a}\left(\begin{matrix}{\ \ }1&-1\\
-1&{\ \ }1
\end{matrix}\right)+\hat \Gamma_{(0)}(z)\,.
$$
For any
$$\psi=\psi_z+G(z)\Pi\,\Gamma_{\Pi,\Theta}(z)^{-1}\Pi\tau\psi_z\in
\D(A_{\Pi,\Theta})\subseteq H^2(0,a)\,,\quad z\in \CO\backslash\RE\,,$$ 
one has
$$
\rho\psi=\Pi\,\Gamma_{\Pi,\Theta}(z)^{-1}\Pi\,\tau\psi_z\,,
$$
i.e.
$$
\rho\psi\in\Ran(\Pi)\,,\quad 
\Theta\rho\psi=\Pi(\tau\psi_z-\Gamma(z)\rho\psi)\,.
$$
Thus
$$
\Pi\tau\psi=\Pi(\tau\psi_z
-\Gamma(z)\rho\psi)=\Theta\rho\psi\,.
$$
Since any $\psi\in
H^2(0,a)$ can be decomposed as
$\psi=(\psi-G(z)\rho\psi)+G(z)\rho\psi$ and
$\psi-G(z)\rho\psi\in\D(A)$, 
a straightforward calculation then gives 
$$
A_{\Pi,\Theta}:\D(A_{\Pi,\Theta})\subseteq L^2(0,a)\to
L^2(0,a)\,,\qquad A_{\Pi,\Theta}\psi=\psi''\,,
$$
$$
\D(A_{\Pi,\Theta})=\left\{\psi\in
H^2(0,a)\,:\,\rho\psi\in\Ran(\Pi)\,,\quad \Pi\tau\psi=\Theta \rho\psi
\right\}\,,
$$
where $(\Pi,\Theta)\in\E(\fh)$, $\fh=\CO^{2}$. 
Thus the case $\Pi=0$ reproduces $A$ itself, the case $\Pi=\uno$ gives
the boundary conditions (here
$\theta_{11},\theta_{22}\in\RE\,,\,\theta_{12}
\in\CO$)
\begin{align*}
&\theta_{11}\,\psi(0+)-\psi'(0+)+\theta_{12}\,\psi(a-)=0\,, 
\\ &\bar \theta_{12}\,\psi(0+)+\theta_{22}\,\psi(a-)+\psi'(a-)=0\,,
\end{align*}
and the case $\Pi=w\otimes w$,
$w\equiv(w_1,w_2)$ 
an unitary
vector in $\CO^2$, gives the boundary conditions (here $\theta\in\RE$)
$$
w_2\,\psi(0+)-w_1\,\psi(a-)=0\,,
$$
$$
\bar w_1\,(\theta\,\psi(0+)-\psi'(0+))
+\bar w_2\,(\theta\,\psi(a-)+\psi'(a-))=0\,.
$$
The resolvent of $A_{\Pi,\Theta}$ is obtained by inserting the above
expressions for $\Gamma(z)$, $G(z)$ and $G(\bar z)^*$ into (\ref{krein}).
\end{example}
\begin{example}(The Laplacian on a bounded graph) 
Let 
$$
A=\oplus_{k=1}^n A_k: \oplus_{k=1}^n \D(A_k)\subseteq\oplus_{k=1}^n
L^2(0,a_k)\to \oplus_{k=1}^n L^2(0,a_k)\,,
$$
$$A_k:\D(A_k)\subseteq L^2(0,a_k)\to
L^2(0,a_k)\,,\qquad A_k\psi=\psi''\,,$$ 
$$\D(A_k)=\{\psi\in H^2(0,a_k)\,:\,\psi(0+)=\psi(a_k-)\}\,,$$
$$
\tau=\oplus_{k=1}^n\tau_k
:\oplus_{k=1}^n H^2(0,a_k)\to\CO^{2n}\,,
$$
$$
\tau_k
:H^2(0,a_k)\to\CO^{n}\,,\quad\tau_k\psi_k,
:=(\psi'_k(0+),-\psi'_{k}(a_k-))\,.
$$
One has 
$$
G(z)=\oplus_{k=1}^n G_k(z):\CO^{2n}\to\oplus_{k=1}^n L^2(0,a_k)
$$
and
$$
G(\bar z)^*=\oplus_{k=1}^n G_k(\bar z)^*:\oplus_{k=1}^n 
L^2(0,a_k)\to\CO^{2n}\,,$$
where $G_k(z)$ and $G_k(\bar z)^*$ are given by (\ref{GZ}) and
(\ref{GZ*}) with $a=a_k$. Analogously 
$$
\Gamma(z)=\oplus_{k=1}^n \Gamma_k(z):\CO^{2n}\to\CO^{2n}\,,
$$
where $\Gamma_k(z)$ is defined as in (\ref{gammaz}) (as in
(\ref{gamma0}) when $z=0$) with $a=a_k$.\par
Proceeding as in the previous example one has
$$
A_{\Pi,\Theta}:\D(A_{\Pi,\Theta})\subseteq \oplus_{k=1}^n 
L^2(0,a_k)\to\oplus_{k=1}^n 
L^2(0,a_k)\,,
$$
$$ A_{\Pi,\Theta}(\psi_1,\dots,\psi_n)=(\psi''_1,\dots,\psi''_n)\,,$$
\begin{align*}
&\D(A_{\Pi,\Theta})\\
=&\left\{\Psi\equiv(\psi_1,\dots,\psi_n)\in\oplus_{k=1}^n
H^2(0,a_k)\,:\,\rho\Psi\in\Ran(\Pi)\,,\quad \Pi\tau\Psi=\Theta \rho\Psi
\right\}\,,
\end{align*}
where $(\Pi,\Theta)\in\E(\fh)$, $\fh=\CO^{2n}$, and
$$
\rho=\oplus_{k=1}^n\rho_k
:\oplus_{k=1}^n H^2(0,a_k)\to\CO^{2n}\,,
$$
$$
\rho_k
:H^2(0,a_k)\to\CO^{n}\,,\quad\rho_k\psi_k
:=(\psi_k(0+),\psi_{k}(a_k-))\,.
$$
Moreover
\begin{align*}
&(-A_{\Pi,\Theta}+z)^{-1}=\oplus_{k=1}^n(-A_{k}+z)^{-1}\\&+
(\oplus_{k=1}^nG_k(z))\Pi\,(\Theta+\Pi(\oplus_{k=1}^n\Gamma_k(z))\Pi)^{-1}
\Pi (\oplus_{k=1}^nG_k(\bar z)^*)\,.
\end{align*}
The self-adjoint operator $A_{\Pi,\Theta}$ describes the Laplacian on
a bounded graph with $n$ edges, the $k$-th edge being identified with the
segment $[0, a_k]$. By a similar construction it is possible to define the
Laplacian on a graph with unbounded external lines. The boundary conditions 
$\Pi\tau\Psi=\Theta \rho\Psi$ specify the connectivity of the
graph. For such a kind of operators, in the case the parametrisation is given
by a couple of $n\times n$ matrices satisfying (\ref{comm}) and 
(\ref{nondeg}), see \cite{[KS]} and see \cite{[AP]} for the corresponding
resolvent formula. In \cite{[Ku]}, Theorem 6, it was shown that such
a parametrization in terms of a couple of matrices can be re-phrased
in a way that coincides with our one (no resolvent formula was
provided there).
\end{example}
\begin{example} (The Laplacian with $n$ point interactions) Let $$
A:H^2(\RE^3)\subseteq L^2(\RE^3)\to L^2(\RE^3)\,,\qquad
A\psi=\Delta\psi\,,$$ 
$$\tau :H^2(\RE^3)\to\CO^n\,,\qquad
\tau\phi\equiv(\psi(y_1),\dots,\psi(y_n))\,,$$
where $y_k\in\RE^3$, $1\le k\le n$. 
Here $H^2(\RE^3)\subset C_b(\RE^3)$ 
denotes the usual Sobolev-Hilbert space of square
integrable functions with square integrable second order
(distributional) partial derivatives. Thus 
$$
S:\D(S)\subset L^2(\RE^3)\to L^2(\RE^3)\,, \quad S\psi=\Delta\psi\,,
$$
$$
\D(S):=\left\{\psi\in H^2(\RE^3)\,:\, \psi(y_k)=0\,, \,1\le k\le n\right\}\,.
$$
Since the kernel of the resolvent of $\Delta$ is given by 
$$(-\Delta+z)^{-1}(x_1,x_1)
=\frac{e^{-\sqrt z\,|x_1-x_2|}}{4\pi|x_1-x_2|}\,,\qquad \text{\rm Re}\sqrt z>0\,,$$ 
one has, if $\zeta\equiv(\zeta_1,\dots,\zeta_n)$,
\begin{equation}\label{G}
G(z):\CO^n\to L^2(\RE^3)\,,\qquad [G(z)\zeta](x)=\sum_{k=1}^n
\frac{e^{-\sqrt z\,|x-y_k|}}{4\pi|x-y_k|}\
\zeta_k
\end{equation}
and
\begin{equation}\label{G*}
G(\bar z)^*:L^2(\RE^3)\to\CO^n\,,\quad G(\bar z)^*\equiv(G(\bar
z)^*_1,\dots, G(\bar z)^*_n)\,, 
\end{equation}
$$
G(\bar z)^*_k\psi:=\int_{\RE^3}\frac{e^{-\sqrt z\,|x-y_k|}}{4\pi|x-y_k|}\,
\psi(x)\,dx\,.
$$
By (\ref{resg}) the $k$-th component of $(z-w)G(\bar w)^*G(z)\zeta$ is 
\begin{align*}
&(z-w)(G(\bar w)^*G(z)\zeta)_k=(\tau(G(w)-G(z))\zeta)_k\\
=&\lim_{x\to y_k}\frac{e^{-\sqrt w\,|x-y_k|}-e^{-\sqrt
    z\,|x-y_k|}}{4\pi|x-y_k|}\ \zeta_k
+\sum_{j\not=k}\left(
\frac{e^{-\sqrt w\,|y_k-y_j|}}{4\pi|y_k-y_j|}
-\frac{e^{-\sqrt z\,|y_k-y_j|}}{4\pi|y_k-y_j|}\right) 
\zeta_j\,
\end{align*}
so that, according to (\ref{gamma}), 
we can take $\Gamma(z):\CO^n\to\CO^n$ to be represented by the 
matrix with
components 
\begin{equation*}
\Gamma_{kj}(z)
=
\begin{cases}\frac{\sqrt
z}{4\pi}&k=j\\
-\frac{e^{-\sqrt z\,
|y_k-y_j|}}{4\pi|y_k-y_j|}& k\not=j\,.\end{cases}
\end{equation*}
Note that we can alternatively define $\Gamma(z)$ by
\begin{equation}\label{altern}
\Gamma(z):=\hat\Theta+\hat\tau(G(0)-G(z))\,,
\end{equation}
where 
\begin{equation}\label{G0}
G(0):\CO^n\to L^2_{loc}(\RE^3)\,,\quad [G(0)\zeta](x):=\sum_{k=1}^n
\frac{\zeta_k}{4\pi|x-y_k|}\,,
\end{equation}
$\hat\tau$ is the extension of $\tau$ to
$H^2_{loc}(\RE^3)\subset C_b(\RE^3)$ and the symmetric operator $\hat\Theta$ is represented by the matrix with
components 
\begin{equation*}
\hat\Theta_{kj}
=
\begin{cases}0&k=j\\
-\frac{1}{4\pi|y_k-y_j|}& k\not=j\end{cases}\,.
\end{equation*}
Given, according to Theorem \ref{estensioni},
$$
\psi=\psi_z+G(z)\zeta_\psi\in D(\Delta_{\Pi,\Theta})\,,\quad
\zeta_\psi:=\Pi\,\Gamma_{\Pi,\Theta}(z)^{-1}\Pi\tau\psi_z\,\quad z\in\CO\backslash\RE\,,
$$
one has, by using (\ref{altern}),
\begin{align*}
&\Pi\hat\tau(\psi-G(0)\zeta_\psi)=\Pi\tau\psi_{z}+\Pi(\hat\Theta-\Gamma(z))\Pi
\zeta_\psi\\
&=\Pi\tau\psi_{z}-\Gamma_{\Pi,\Theta}(z)\zeta_{\psi}+
\Pi\hat\Theta\Pi\zeta_\psi+\Theta\zeta_\psi\,.
\end{align*}
This gives $\Pi\hat\tau_0\psi=\Theta\,\zeta_\psi$,
where the renormalised trace operator $\hat\tau_0$ is defined by 
\begin{equation}\label{tau}
\hat\tau_0:\D(\Delta_{\Pi,\Theta})\to\CO^n\,,\quad
\hat\tau_0\psi:=\hat\tau(\psi-G(0)\zeta_\psi)-\hat\Theta\zeta_\psi\,.
\end{equation}
Note that such a definition says that 
the $k$-th component of $\hat\tau_0\psi$ is given by
$$
(\hat\tau_0\psi)_k=\lim_{x\to y_k}\left(\psi(x)-\frac{1}{4\pi}\,
\frac{\zeta_{k}}{|x-y_k|}\right)\,,\quad 1\le k\le n\,,
$$
where here $\zeta_k$ denotes the $k$-th component of $\zeta_\psi$.
By Theorem \ref{estensioni}
one has
\begin{align*}
&\Delta_{\Pi,\Theta}\psi=\Delta\psi_z+zG(z)\zeta_\psi\\
=&\Delta(\psi-
G(0)\zeta_\psi)+zG(z)\zeta_\psi-\Delta(G(z)-G(0))\zeta_\psi\\
=&
\Delta(\psi-G(0)\zeta_\psi)\,.
\end{align*}
In conclusion, for any $(\Pi,\Theta)\in\E(\fh)$, $\fh=\CO^n$, one has
$$
\Delta_{\Pi,\Theta}:\D(\Delta_{\Pi,\Theta})\subseteq L^2(\RE^3)\to
L^2(\RE^3)\,,\qquad\Delta_{\Pi,\Theta}\psi:=\Delta\psi_0\,,
$$
\begin{align*}
&\D(\Delta_{\Pi,\Theta}):=\{\psi\in
L^2(\RE^3)\,:\,\psi=\psi_0+G(0)\zeta_\psi\,,\\& \psi_0\in
H^2_{loc}(\RE^3)\,,\quad\zeta_\psi\in\Ran(\Pi)\,,\quad\Pi\hat\tau_0\psi
=\Theta\,\zeta_\psi\}
\end{align*}
and 
$$
(-\Delta_{\Pi,\Theta}+z)^{-1}=(-\Delta+z)^{-1}+
G(z)\Pi\,(\Theta+\Pi\,\Gamma(z)\Pi)^{-1}\Pi G(\bar z)^*\,.
$$
The case $\Pi=\uno$, $\Theta$ diagonal, 
reproduces the self-adjoint extensions appearing
in \cite{[AGHH]} and references therein. For the
general case (when the parametrisation is given
by a couple of $n\times n$ matrices satisfying (\ref{comm}) and 
(\ref{nondeg})) see \cite{[Pa]}.  
\end{example} 
\begin{example} (The Laplacian with $n$ point interactions, 
the vector-valued case) The previous example can be generalised by taking
$$
A:H^2(\RE^3;\ff)\subseteq L^2(\RE^3;\ff)\to L^2(\RE^3;\ff)\,,$$
\begin{equation}\label{A}
A\psi:=\Delta\psi+B\psi\,,\end{equation} 
where $B$ is a symmetric operator in the $d$-dimensional Hilbert space
$\ff$, and
$$\tau :H^2(\RE^3;\ff)\to\fh\,,\qquad \fh=\oplus_{k=1}^n\ff\,,\quad
\tau\phi\equiv(\psi(y_1),\dots,\psi(y_n))\,,.$$ 
By using the unitary isomorphisms 
$$\ff\simeq\CO^d\,,\quad
\oplus_{k=1}^n\ff\simeq\oplus_{i=1}^d\CO^n\,,\quad 
L^2(\RE^3;\ff)\simeq\oplus_{i=1}^dL^2(\RE^3)$$ 
induced by the
orthonormal basis $\e_1,\dots, \e_d$ made of the normalised eigenvectors
of $B$, and denoting by $b_1,\dots,b_d$ the corresponding eigenvalues,
one has
$$
G(z)=\oplus_{i=1}^d G_i(z):\oplus_{i=1}^d \CO^n\to \oplus_{i=1}^d L^2(\RE^3)\,,
$$
$$
G(\bar z)^*=\oplus_{i=1}^d G_i(\bar z)^*:\oplus_{i=1}^d L^2(\RE^3)\to
\oplus_{i=1}^d \CO^n\,.
$$
Here
$$
G_i(z):\CO^n\to L^2(\RE^3)
$$
and
$$
G_i(\bar z)^*:L^2(\RE^3)\to \CO^n 
$$
are defined by evaluating the operators $G(\cdot)$ and $G({
\,\bar\cdot}\,)^*$ given in 
(\ref{G}) and (\ref{G*}) at $z-b_i$. Analogously
$$\Gamma(z)=\oplus_{i=1}^d\Gamma_i(z):\oplus_{i=1}^d\CO^n\to\oplus_{i=1}^d
\CO^n\,,$$ 
where $\Gamma_i(z)$ is represented by the matrix with
components 
\begin{equation*}
\Gamma_{i,kj}(z)
=
\begin{cases}\frac{\sqrt
{z-b_i}}{4\pi}&k=j\\
-\frac{e^{-\sqrt {z-b_i}\,
|y_k-y_j|}}{4\pi|y_k-y_j|}& k\not=j\,.\end{cases}
\end{equation*}
Proceeding as in the previous example one has,
for any $(\Pi,\Theta)\in\E(\fh)$, $\fh=\oplus_{i=1}^n\ff
\simeq\oplus_{i=1}^d\CO^n\simeq\CO^{nd}$,
$$
A_{\Pi,\Theta}:\D(A_{\Pi,\Theta})\subseteq L^2(\RE^3;\ff)\to
L^2(\RE^3;\ff)\,,\qquad A_{\Pi,\Theta}\psi:=\Delta\psi_0+B\psi_0\,,
$$
\begin{align*}
&\D(A_{\Pi,\Theta}):=\{\psi\in
L^2(\RE^3;\ff)\,:\,\psi=\psi_0+(\oplus_{i=1}^d G(0))\,\zeta_\psi\,,\\& 
\psi_0\in H^2_{loc}(\RE^3;\ff)\,,\quad\zeta_\psi\in\Ran(\Pi)\,,
\quad\Pi\hat\tau_0\psi=\Theta\,\zeta_\psi\}\,,
\end{align*}
\begin{align*}
&(-A_{\Pi,\Theta}+z)^{-1}=(-A+z)^{-1}\\&+
(\oplus_{i=1}^dG_i(z))\Pi\,(\Theta+\Pi(\oplus_{i=1}^d\Gamma_i(z))\Pi)^{-1}
\Pi (\oplus_{i=1}^dG_i(\bar z)^*)\,,
\end{align*}
where $G(0)$ is defined in (\ref{G0}) and $\hat\tau_0$ is here defined
component-wise through (\ref{tau}) by writing 
$\psi=\psi_1\,\e_1+\dots+\psi_d\,\e_d$.\par
By the unitary isomorphism $L^2(\RE^3)\otimes
\ff\simeq L^2(\RE^3;\ff)$ given by $\psi\otimes\zeta\mapsto \psi\zeta$, 
which transforms $\Delta\otimes\uno+\uno\otimes B$ into $A$ defined in 
(\ref{A}), and by taking
$\ff=\otimes_{k=1}^n\CO^2\simeq \CO^{2^n}$, 
this example reproduces (for a
particular choice of $B$) the self-adjoint extensions given in
\cite{[CCF]} describing systems made of a spin-less quantum particle
and an array of $n$ spin $1/2$ (there the parametrisation is given
by a couple of $n2^n\times n2^n$ matrices 
satisfying (\ref{comm}) and 
(\ref{nondeg})).
\end{example}
\begin{example}(The Laplacian on a bounded domain) This last example is the
  extension of Example 5.1 to $n$ dimensions. The main difference is
  due to the infinite dimensionality of $\fh$, i.e. the defect indices
  of $S$ are infinite in this case. 
This requires the use of some not trivial analytic results 
which we entirely take from \cite{[G1]} and \cite{[LM]}. 
However, apart from these technical issues, we follow
  the same path as in the much simpler Example 5.1. This
  leads to the reproduction of the results obtained (for 
general strongly elliptic operators) in \cite{[G1]}, Chapter III, about
  the complete classification in terms of boundary conditions 
of the self-adjoint extensions of the
  minimal Laplacian on a bounded domain. The study of boundary value
  problems by means of self-adjoint extensions goes back to
  \cite{[C]} and was further developed in \cite{[V]}.\par 
Given $\Omega\subset\RE^n$, $n>1$, 
a bounded open set with a boundary $\partial\Omega$ which is a
smooth embedded sub-manifold 
(these hypotheses could
be weakened), $H^m(\Omega)$ denotes the usual Sobolev-Hilbert
space of functions on $\Omega$ with square integrable partial
(distributional) 
derivatives of any order $k\le m$ and $H^s(\partial\Omega)$,
$s$ real, denotes the fractional order Sobolev-Hilbert space defined, 
since here $\partial\Omega$ can be made a smooth compact Riemannian
manifold, as the
completion of $C^\infty(\partial\Omega)$ with respect of the scalar
product 
$$
\langle f,g\rangle_{H^s(\partial\Omega)}
:=\langle f,(-\Delta_{LB}+1)^{s}g\rangle_
{L^2(\partial\Omega)}\,.
$$ 
Here the self-adjoint
operator $\Delta_{LB}$ is the Laplace-Beltrami operator in 
$L^2(\partial\Omega)$. With such a definition $(-\Delta_{LB}+1)^{s/2}$ 
can be extended to a unitary map, 
which we denote by the same
symbol, 
$$(-\Delta_{LB}+1)^{s/2}:H^{r}(\partial\Omega)
\to H^{r-s}(\partial\Omega)\,.
$$
For successive notational convenience we pose
$$
\Lambda:=(-\Delta_{LB}+1)^{1/2}:H^{s}(\partial\Omega)
\to H^{s-1}(\partial\Omega)\,,\quad \Sigma:=\Lambda^{-1}\,.
$$
The continuous and surjective
linear operator 
$$
\gamma:H^2(\Omega)\to
H^{3/2}(\partial\Omega)\oplus
H^{1/2}(\partial\Omega)\,,\qquad \gamma\phi:=(\rho\phi,\tau\phi)\,,
\,,
$$
is defined (see e.g. \cite{[LM]}, Chapter 1, Section 8.2) as the unique 
bounded extension of 
$$
\tilde \gamma:C^{\infty}(\bar\Omega)\to C^{\infty}(\partial\Omega)\times C^{\infty}(\partial\Omega)
\,,\qquad \tilde \gamma\phi:=(\tilde\rho\phi,\tilde\tau\phi)\,,
$$
where
$$
\tilde\rho\phi\,(x):=\phi\,(x)\,,\quad
\tilde\tau\phi\,(x):=n(x){}^{}\!\cdot\!\nabla\phi\,(x)
\equiv\frac{\partial
  \phi}{\partial n}\,(x)\,,\quad x\in\partial\Omega\,,
$$
and $n$ denotes the inner normal vector on $\partial\Omega$. 
By Green's formula the linear
operator $\gamma$  can be further extended
(see \cite{[LM]}, Chapter 2, Section 6.5) to a 
continuous map
$$
\hat\gamma:\D(\Delta_{max})\to H^{-1/2}(\partial\Omega)\oplus 
H^{-3/2}(\partial\Omega)\,,\qquad \hat \gamma\phi
=(\hat\rho\phi,\hat\tau\phi)\,,
$$
where
$$
\D(\Delta_{max}):=\left\{\phi\in L^2(\Omega)\,:\, \Delta\phi\in L^2(\Omega)\right\}\,.
$$ 
Let
us remark that $H^2(\Omega)$ is strictly contained in
$\D(\Delta_{max})$ when $n>1$; by elliptic regularity one has
(see \cite{[G1]}, Proposition III 5.2, \cite{[LM]}, Chapter 2, Section 7.3)
\begin{align}\label{regularity}
H^2(\Omega)=&\left\{\phi\in\D(\Delta_{max})\,:\,\hat\rho\phi\in H^{3/2}
(\partial\Omega)\right\}\\=&
\left\{\phi\in\D(\Delta_{max})\,:\,\hat\tau\phi\in H^{1/2}
(\partial\Omega)\right\}\,.
\end{align}
Now let $A$ be the self-adjoint operator in $L^2(\Omega)$ given by the
Dirichlet Laplacian
$$
A:\D(A)\subseteq L^2(\Omega)\to L^2(\Omega)\,\quad A\psi\equiv
\Delta^D\psi=\Delta\psi\,,
$$
$$
\D(\Delta^D)=\left\{\psi\in H^2(\Omega)\,:\, \rho\psi=0\right\}
$$
and let $\tau:\H_A\to\fh$, with $\fh=H^{1/2}(\partial\Omega)$, be 
the normal derivative operator along $\partial\Omega$ defined
above. Thus we are looking for all self-adjoint extensions of the
symmetric operator given by the minimal Laplacian 
$$S:\D(S)\subseteq L^2(\Omega)\to
L^2(\Omega)\,,\qquad S\psi\equiv\Delta_{min}\psi:=\Delta\psi\,,$$
\begin{align*}
\D(S)\equiv\D(\Delta_{min})\equiv H^2_0(\Omega)
:=\left\{\psi\in
H^2(\Omega)\,:\,\rho\psi=\tau\psi=0\right\}\,.
\end{align*}
Note that by defining the maximal Laplacian
$\Delta_{max}$ as the distributional Laplacian
restricted to $\D(\Delta_{max})$,
one has $\Delta_{max}=(\Delta_{min})^*$.\par
By the definition of $G\left(0\right)$ one has, for any $h\in
  H^{1}(\partial\Omega)$ and for any $\psi\in L^2(\Omega)$,  
$$
\langle G(0)h,\psi\rangle_{L^2(\Omega)}=
-\langle
\Lambda h,\tau(\Delta^D)^{-1}\psi
\rangle_{L^2(\partial\Omega)}\,.
$$
Therefore
$$[G(0)h](x)=-\int_{\partial\Omega}\Lambda h(y)\,\frac{\partial
  }{\partial n}\,g(x,y)\, d\sigma(y)\,,$$
where $g$ is the Dirichlet Green function of $\Omega$ for the
Laplacian, and so 
$$
G(0)=K\Lambda\,,
$$
where $K$ denotes the Poisson operator, i.e. $K:H^{-1/2}(\partial\Omega)\to \D(\Delta_{max})$ is the continuous linear
operator (see e.g. \cite{[LM]}, Chapter 2, Section 6) 
which solves 
the  
Dirichlet boundary value problem
\begin{align}\label{bvp}
&\Delta Kh=0\,,\\
&\hat\rho\, Kh=h\,.
\end{align} 
By (\ref{resg}) one has
\begin{equation}\label{giz}
G(z)=G(0)-z(-\Delta^D+z)^{-1}G(0)=-\Delta^D(-\Delta^D+z)^{-1}K\Lambda
\end{equation}
and so $G(z)h\in \D(\Delta_{max})$, $h\in H^{1/2}(\partial\Omega)$, 
solves the Dirichlet boundary value problem
\begin{align*}
&\Delta G(z)h=zG(z)h\,,\\
&\hat\rho\, G(z)h=\Lambda h\,.
\end{align*}
Thus $$\Ran(G(z))\cap\D(\Delta^D)=\{0\}$$
and condition (\ref{inj}) is satisfied.\par 
Now, according to (\ref{gammahat}), we define the bounded linear operator   
$$
\Gamma(z):
H^{1/2}(\partial\Omega)\to H^{1/2}(\partial\Omega)\,,
$$
\begin{equation}\label{gamz} 
\Gamma(z):= \hat\Gamma_{(0)}(z)\equiv\tau(G(0)- G(z))
=z\,\tau(-\Delta^D+z)^{-1}K\Lambda\,.
\end{equation}
Since $\D(\Delta^D_{\Pi,\Theta})\subseteq
\D(\Delta_{max})$, the trace operators $\hat\rho$ and
$\hat\tau$ act on $\D(\Delta^D_{\Pi,\Theta})$. Thus for any 
$\psi\in\D(\Delta^D_{\Pi,\Theta})$, which according to
Theorem \ref{estensioni} can be written as
$$\psi=\psi_z+G(z)\Pi\,\Gamma_{\Pi,\Theta}(z)^{-1}\Pi\tau\psi_z\in
\,,\quad z\in \CO\backslash\RE\,,$$ 
one has
\begin{equation}\label{rhohat}
\hat\rho\psi=\Lambda \Pi\,\Gamma_{\Pi,\Theta}(z)^{-1}\Pi\,\tau\psi_z\,,
\end{equation}
i.e.
\begin{equation}\label{rhohatbis}
\Sigma\hat\rho\psi\in\D(\Theta)\subseteq\Ran(\Pi)\,,\quad
\Theta \Sigma\hat\rho\psi
=\Pi(\tau\psi_z-\Gamma(z)\Sigma\hat\rho\psi)\,\,,
\end{equation}
and
\begin{equation}\label{tauhat}
\hat\tau\psi=\tau\psi_z
-\Gamma(z)\Sigma\hat\rho\psi
+\hat\tau G(0)\Sigma\hat\rho\psi\,.
\end{equation}
Such relations (\ref{rhohat})-(\ref{tauhat}) show that for any
$\psi\in\D(\Delta^D_{\Pi,\Theta})$
the regularised trace operator $\hat\tau_0$ defined by 
$$\hat\tau_0\psi:=\hat\tau(\psi-G(0)\Sigma\hat\rho\psi)$$
is  $H^{1/2}(\partial\Omega)$-valued and the boundary condition
$$\Pi\hat\tau_0\psi=\Theta \Sigma\hat\rho\psi$$ holds true. 
By elliptic regularity one can define $\hat\tau_0$ on a larger domain: 
by (\ref{regularity}) one has
$$\D(\Delta_{max})\cap\Ker(\hat\rho)=\D(\Delta^D)$$
and so, for any $\psi \in \D(\Delta_{max})$,  
$\psi-G(z)\Sigma\hat\rho\psi$ 
belongs to $\D(\Delta^{D})$. Thus (see 
\cite{[G1]}, Theorem III 1.2)
$$
\hat\tau_0:\D(\Delta_{max})\to H^{1/2}(\partial\Omega)\,,
$$
$$
\hat\tau_0\psi=\tau(\psi-G(0)\Sigma\hat\rho\psi)
=\tau(\psi-K\hat\rho\psi)=\hat \tau\psi-P\hat\rho\psi\,,
$$
where the bounded linear operator $P$, 
known as the Dirichlet-to-Neumann operator over
$\partial\Omega$, is given by (see e.g. \cite{[G1]}, Theorem III 1.1)
$$
P:H^{-1/2}(\partial\Omega)\to 
H^{-3/2}(\partial\Omega)\,,\quad
P:=\hat\tau\, K\,.
$$ 
Since
$$
\psi-G(0)\Sigma\hat\rho\psi
-\left(\Delta^D\right)^{-1}\Delta\psi\in\Ker(\Delta^D)=\{0\}\,,
$$
alternatively one can define $\hat\tau_0$ by 
(see \cite{[G1]}, Theorem III 1.2)
$$
\hat\tau_0:\D(\Delta_{max})\to H^{1/2}(\partial\Omega)\,,\qquad
\hat\tau_0:=\tau\left(\Delta^D\right)^{-1}\Delta\,.
$$
Since any $\psi\in \D(\Delta_{max})$ can be decomposed as
\begin{equation*}
\psi=(\psi-G(z)\Sigma\hat\rho\psi)
+G(z)\Sigma\hat\rho\psi
\end{equation*}
and
\begin{align*}
\Delta^D_{\Pi,\Theta}\psi=&\Delta^D\psi_z
+z\,G(z)\Sigma\hat\rho\psi\\
=&\Delta\psi+(-\Delta+z)G(z)\Sigma\hat\rho\psi\\
=&\Delta\psi\,,
\end{align*}
in conclusion  by Theorem \ref{estensioni} one obtains, for any $(\Pi,\Theta)\in\E(\fh)$, $\fh=H^{1/2}(\partial\Omega)$,
$$
\Delta^D_{\Pi,\Theta}:\D(\Delta^D_{\Pi,\Theta})\subseteq L^2(\Omega)\to
L^2(\Omega)\,,\qquad \Delta^D_{\Pi,\Theta}\psi=\Delta\psi\,,
$$
\begin{align*}
\D(\Delta^D_{\Pi,\Theta})
=\left\{\psi\in
\D(\Delta_{max})\,:\,\Sigma\hat\rho\psi\in\D(\Theta)\,,
\quad\Pi\hat\tau_0\psi=\Theta \Sigma\hat\rho\psi
\right\}
\end{align*}
and
\begin{equation*}
(-\Delta^D_{\Pi,\Theta}+z)^{-1}
=(-\Delta^D+z)^{-1}+
G(z)\Pi\,(\Theta+\Pi\,\Gamma(z)\Pi)^{-1}\Pi G(\bar z)^*\,,
\end{equation*}
where
$G(z)$ and $\Gamma(z)$ are given in (\ref{giz}) and (\ref{gamz}) 
respectively.\par
Since $\Sigma:H^{-1/2}(\partial\Omega)\to H^{1/2}(\partial\Omega)$
is unitary and $\Pi$ is an orthogonal projection in
$H^{1/2}(\partial\Omega)$, 
we can re-parametrise the extensions $\Delta^D_{\Pi,\Theta}$
by the couple $(X,L)$,
where $X\subseteq H^{-1/2}(\partial\Omega)$ is a closed subspace and
$L$ is self-adjoint from $X$ to its dual (with respect to the 
$L^2(\partial\Omega)$ pairing) $X^*$. In this
case one has
\begin{align*}
\D(\Delta^D_{X,L})
=\left\{\psi\in
\D(\Delta_{max})\,:\,\hat\rho\psi\in\D(L)\,,
\quad \left.\hat\tau_0\psi\right|_X=L\hat\rho\psi
\right\}\,,
\end{align*}
where the boundary condition 
$\left.\hat\tau_0\psi\right|_X=L\hat\rho\psi$ means
$$
\forall\,f\in X\cap L^2(\partial\Omega)\,,\quad
\langle\hat\tau_0\psi,f\rangle_{L^2(\partial\Omega)}=
\langle L\hat\rho\psi,f\rangle_{L^2(\partial\Omega)}\,.
$$
This alternative description reproduces the one obtained (for a
general strongly elliptic operator) in \cite{[G1]}, Theorem III 4.1 
(also see \cite{[G2]}, \cite{[G3]} and references therein). \par
The usual Robin-like boundary 
conditions can be recovered in the following way:
since $G(0)h$ solves (\ref{bvp})-(5.14), if $h\in
H^{5/2}(\partial\Omega)$ 
then $G(0)h\in H^2(\Omega)$ by elliptic regularity (see
(\ref{regularity})) and so we can define
the unbounded operator in $H^{1/2}(\partial\Omega)$
$$
\Theta_0:H^{5/2}(\partial\Omega)\subseteq H^{1/2}(\partial\Omega)\to 
H^{1/2}(\partial\Omega)\,,\quad$$
$$ \Theta_0:=-\tau G(0)=-\tau K\Lambda=-P\Lambda\,.
$$ 
$\Theta_0$ is symmetric since $P$ is $L^2(\partial\Omega)$-symmetric
by Green's formula and so, 
given any $L^2(\partial\Omega)$-symmetric bounded linear operator  
$$B:H^{3/2}(\partial\Omega)\to
H^{1/2}(\partial\Omega)\,,
$$
we can define the unbounded symmetric operator
$$
\Theta_B:H^{5/2}(\partial\Omega)\subseteq H^{1/2}(\partial\Omega)\to 
H^{1/2}(\partial\Omega)\,,$$
$$\Theta_B:=\Theta_0+B\Lambda=\left(-P+B\right)\Lambda\,.
$$
Since $\Lambda^2$ is self-adjoint and $(-P+B)\Sigma$ is bounded, one
has $$\Theta_B^*=\Lambda^2((-P+B)\Sigma)^*\supseteq \Theta_B\,.$$ Thus 
$((-P+B)\Sigma)^*$ coincides with $\Sigma^2(-P+B)\Lambda$ on 
$H^{5/2}(\partial\Omega)$ and therefore 
$\Theta_B$ is self-adjoint if and only if
\begin{equation}\label{iff}
\{\phi\in H^{-1/2}(\partial\Omega)\,: 
(-P+B)\phi\in H^{1/2}(\partial\Omega)\}\subseteq H^{3/2}(\partial\Omega)\,.
\end{equation}
Here, by a slight abuse of notation, 
we used the same symbol $B$ to denote the bounded extension of $B$
given by the operator on 
$H^{-1/2}(\partial\Omega)$ to $H^{-3/2}(\partial\Omega)$ obtained by
considering its 
adjoint (with respect to the $L^2(\partial\Omega)$ pairing).
By elliptic regularity one has (see \cite{[G1]}, Theorem III 5.4)
$$
\{\phi\in H^{-1/2}(\partial\Omega)\,: 
P\phi\in H^{s}(\partial\Omega)\}\subseteq H^{s+1}(\partial\Omega)\,.
$$
Thus (\ref{iff}) holds true (by an
iterative argument) when $B$ maps 
$H^{s}(\partial\Omega)$ to
$H^{s-1+\epsilon}(\partial\Omega)$ for any 
$s\in\left[-\frac{1}{2},\frac{3}{2}\right]$ and for some
$\epsilon>0$ (see \cite{[G1]}, Chapter III, Section 6, for a similar
kind of results). 
So $B$ could be a pseudo-differential operator of
order $1-\epsilon$, in particular the multiplication by a (sufficiently
regular) function. 
The case in which $\epsilon=0$ is more delicate and a direct analysis is
required: (\ref{iff}) holds true when $-P+B$ is an
elliptic  pseudo-differential operator of order one; this can be checked
by studying its symbol (see \cite{[G3]}, Corollary 9.34, 
for the case in which $A=-\Delta+1$, $B$ is differential
  of order one and
  $\Omega=\RE^n_+$).
\par 
By taking $\Pi=\uno$, 
$\Theta=\Theta_B$, one then obtains the self-adjoint extension
$$
\Delta_{B}:\D(\Delta_{B})\subseteq L^2(\Omega)\to
L^2(\Omega)\,,\qquad \Delta_{B}\psi:=\Delta\psi\,,
$$
with domain (the second expression being consequence of elliptic regularity, see 
(\ref{regularity}))
\begin{align*}
\D(\Delta_{B})
:=&\left\{\psi\in
\D(\Delta_{max})\,:\,\hat\rho\psi\in H^{3/2}(\partial\Omega)\,,
\, \hat\tau\psi=B\hat\rho\psi
\right\}\\
\equiv&\left\{\psi\in
H^2(\Omega)\,:\,\tau\psi=B\rho\psi
\right\}\,.
\end{align*}
We refer to \cite{[G1]}, Chapter III, Section 6, for a 
detailed study of the properties of $=-P+B$ and their relations
with properties of $\Delta_B$.

\end{example}

\end{section}
\vskip10pt\noindent
{\bf Acknowledgement. }We thank Gerd Grubb for valuable suggestions regarding
  Example 5.5, Mark Malamud and Konstantin Pankrashkin for 
useful bibliographic remarks.

\end{document}